# Hands-on learning at a world-class telescope

**Elisabeta Lusso[1,2], Lapo Casetti[1,2,3], Maurizio Pancrazzi[4], Marco Romoli[1,2]**

[1]Dipartimento di Fisica e Astronomia, Università di Firenze, via G. Sansone 1, 50019 Sesto Fiorentino, Firenze, Italy, [2]INAF–Osservatorio Astrofisico di Arcetri, Largo Enrico Fermi 5, I-50125 Firenze, Italy, [3]INFN–Sezione di Firenze, via G. Sansone 1, I-50019 Sesto Fiorentino, Italy.[4]INAF–Astrophysical Observatory of Torino, Torino, Italy.

*Abstract*

*For the first time, an Italian University has the possibility to perform a multi-year observing campaign at a world-class telescope. This hands-on experience had a significant impact on the students' university path: from learning specific observing techniques on-site to teamwork and collaboration. In this paper we present the results of an observing campaign carried out at the Telescopio Nazionale Galileo (TNG) located in La Palma (Canary Islands, Spain) by undergraduate students of the Department of Physics and Astronomy, at the University of Firenze.*

# 1. Introduction

The hill of Arcetri in the outskirts of the city of Firenze, where Galileo spent his last years, has represented the pivotal site of research in physics and astronomy in Firenze, hosting both the Institute of Physics of the local University and the Arcetri Astrophysical Observatory, and being for this reason promoted to a historical site of the European Physical Society. During the 1980s, the Department of Astronomy and Space Science of the University was constituted, based in Arcetri alongside with the Physics Department, heir to the Institute of Physics (Mazzoni 2014). The two university departments eventually merged into the Department of Physics and Astronomy in 2010. The astrophysics researchers of the university departments have maintained a close collaboration with the Arcetri Observatory throughout the years. Within this collaboration, in 2017, a private donation to the Arcetri Observatory was partly devolved to the University. This private funding allowed university students, for the first time, to perform an observing campaign at a world-class telescope within a part of the teaching course *Complementi di Astronomia,* which is the subject of the present manuscript. Complementi di Astronomia is offered to bachelor and master students and gives a first introduction to experimental astronomy. Specifically, the course presents the relevant instruments, detectors, and telescope techniques, primarily for observations at optical wavelengths. A substantial part of the course is dedicated to hands-on observing sessions. The main scientific aim of the latter is to determine the angular diameter and the redshift of a sample of local spiral galaxies to provide an estimate of their distances and, possibly, of the Hubble parameter, $H_0$, which describes the current expansion rate of the Universe. Until 2017, observations had been performed at the 152 cm Cassini telescope located in Loiano (Bologna), Italy. In 2017, the students carried out the first international observing campaign at the *Telescopio Nazionale Galileo* (TNG) located at the Roche de los Muchachos in La Palma (Canary Islands, Spain). Such a campaign requires not only a significant budget (not affordable without external funding given the overall reduction in funding from the Italian State to the Universities), but also a considerable logistic effort between the teachers and the staff at the TNG. During the trip, we take the opportunity to visit most of the astronomical telescopes at the Roque.

In this manuscript, we critically discuss our *outside the classroom* experience from different perspectives (e.g., students teamwork on-site and in the lab) and how everything changed after the COVID-19 outbreak. We note that the discussion of the practical part of the lecture will be mostly qualitative. We did not carry out much literature research in first place for two main reasons. First and foremost, the experience we describe is virtually a *unicum* in Italy and, since it is subject to external funding, we did not have the perspective of a more structured teaching path related to on-site observations. This may change in the following years if we can secure funding that allows us to schedule several observing runs during the academic year. Second, we can only now begin to study the statistics of the students attending the experimental part, as it started only 5 years ago and with a 3 year stop due to the pandemic. Further literature review and a follow-up study on how this kind of visit may contribute to the education of students will be carried out in the following academic years.

## 2. Scope of the observations

The TNG is the largest (with a 3.58 m diameter primary mirror) optical telescope of the Italian astronomical community and is managed by the Fundación Galileo Galilei (FGG, Fundación Canaria) on behalf of the Italian National Institute for Astrophysics (INAF)[1]. DOLoReS (Device Optimised for the Low RESolution) is an imager and a low-resolution spectrograph permanently installed at the TNG equipped with photometric filters and grisms. The DOLORES detector consists of one E2V back-illuminated and thinned 2048 × 2048 pixel CCD, with a scale of 0.275 arcsec pixel$^{-1}$, resulting in a field of view of 9.4 × 9.4 arcmin$^2$. The students used this instrument to perform both imaging and low-resolution spectroscopic observations of nearby spiral galaxies ($z<0.2$). Different dispersing elements (grisms) and slit widths ranging from 1″ to 1″.5 were used to perform spectroscopy, securing a spectral coverage in the 3500-8200 Å band with dispersions between 1 and 2.5 Å pixel$^{-1}$. The main scientific aim of the observations is to measure the apparent diameter (from imaging) and redshift (from spectroscopy) of local spiral galaxies to constrain the universe expansion, parameterized by the Hubble parameter, $H_0$. Ancillary science goals are the determination of the star formation rate and the estimate of the mass of the central supermassive black hole. The latter is possible whenever the galaxy observed has an *active galactic nucleus* (AGN hereafter). In an AGN, the supermassive black hole located in the galaxy center is accreting material from its immediate surroundings in a form of a disc, converting the gravitational potential of this matter into heat and, eventually, radiating this energy away. The nucleus of these galaxies shines bright at optical-UV wavelengths and, by computing the full-width at half maximum (FWHM) of specific broad emission lines and the continuum at a given optical wavelength, the students can obtain an estimate of the mass of the black hole. Below we first detail how the students prepare the observing run and the data analysis performed by the students once they are back in the laboratory at their home institution. During the course the students will select the observing targets, perform the observation campaign, reduce, and analyse the collected data.

### *2.1. Preparing the observations*

The students are first preliminarily instructed on how they should use the online archives and what are the constraints to select the objects to create a suitable target list. Namely, the students are asked to select a set of galaxies (usually between 4 and 10 objects) that fulfil a series of criteria ranging from the type of object (i.e., spirals) to its visibility at the telescope (e.g., coordinates, maximum air mass, relative distance from the Moon depending on the night conditions). The students also check whether there is an available spectrum in the literature to verify that bright emission lines (required for the redshift determination) are present. The students are divided in groups of maximum 3 people and each group explores a different interval of coordinates (i.e., the sky visible from the observing site) such that there are no overlaps in the possible targets the students can find. Each group then presents a list of galaxies with the finding charts (i.e., the image of the galaxy within the field-of-view of the telescope), the available magnitudes and surface brightness values at several wavelengths (e.g., *B*, *V* and *R*), the optical spectra from the literature and the integration times for both imaging and spectroscopy with the relative signal-to-noise ratio (*S/N*). Each group delivers a preliminary target list which is then verified by the teachers and merged to create the final one. This is the first step where the students put into practice what they have learnt during the lectures, thus dealing with several data archives of celestial objects, understanding the different nomenclatures, and working with astronomical software tools.

---

[1] http://www.tng.iac.es/

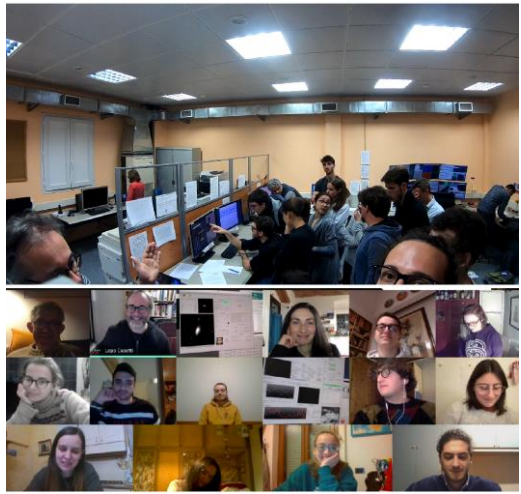

*Figure 1. Top: Students in the control room at the TNG during the last observing run (April 2019) before the COVID-19 outbreak. Bottom: Set-up for the virtual observing. The third window in the first row shows the target observed, whilst the fourth window in the second row shows the monitor with the weather conditions of the night.*

### *2.2. Observing at the telescope (before COVID-19)*

Once the target list is prepared, it is sent to the support astronomer a few days before the observing night for final approval. In the control room, the students are guided by the support astronomer through the several steps to carry out both imaging and spectroscopy, meaning that the students themselves have the responsibility to input the command lines to acquire the target, check whether the source in the field-of-view is correct, select the right filter (or slit) for the imaging (and spectroscopic) observation. The students also need to interface with the telescope operator, who is ultimately responsible for moving the telescope (Figure 1, top panel). Although the students always have the guidance of the support astronomer and the teachers are present during the entire run, they are required to do most of the steps themselves, to *learn and act fast*, exploiting every minute of the roughly three-hour observing time. In fact, a mistake in typing a command line on the terminal, for instance, means that precious observing time is lost, and this could undermine the entire observing schedule. All the steps done during the observations are, by all means, what is customarily done by professional astronomers for their science research. We asked the students, once finished, how the experience had been, and all the students agreed that this was the most stressful part of the entire process. Yet, they felt very rewarded once they knew they got their own data.

*2.3. Data analysis*

Back to the computer laboratory at home, the students must reduce and analyse the data on their own. During this part of the course the students gain sound knowledge and understanding of data handling. Students also become acquainted with several forms of data representation and explore different techniques of regression analysis to determine the 'best-fit' and data modelling. Astronomical data are stored in a FITS format; thus, the students further learn how to visualise them interactively with SAO Image DS9, which is one of the standard tools to examine astronomical images. Students are split again into groups of max 3 people and all the different groups choose one galaxy to analyse. They are provided with a data processing manual describing the main steps to follow for both the imaging and the spectroscopic reduction. They first need to tackle the estimate of the seeing, a parameter that represents the sharpness of a telescope image that depends upon the turbulence of the Earth's atmosphere. Roque de los Muchachos is a very good observing site, with an average seeing of 0.9 arcsec (measured from 2011 May to 2012 February, Gurtubai et al. 2013). The students then start to handle the calibration files, namely bias and flat field, for the data reduction of the images, and analyse their error statistics and readout noise. They create a master bias and flat field by computing the median of the different images (taking into account and removing the possible signature of cosmic rays), which are subsequently applied to the raw data to generate the final calibrated image. An example of raw and reduced images is shown in Figure 2 for the galaxy NGC 3294 observed on March 21$^{st}$, 2021, remotely. The next step is the reduction of the spectroscopic data. The students calibrate the 2D spectral images for the bias, flat-field and correct for the cosmic rays. The 1D spectrum is then extracted together with the one of the background. Helium-Argon lamps are used for the wavelength calibration of both the source and the background spectra, which are interpolated across a common wavelength range. The 1D background spectrum is then subtracted from the source one. The spectrum of the galaxy is finally flux calibrated and corrected for atmospheric absorption by making use of the spectrum of a spectrophotometric standard star observed during the observation run with the same spectral set up. An example of raw and reduced 2D spectral images by employing the reduction pipeline built by a group of students for the galaxy NGC 3294 is presented in Figure 3. The data reduction tasks described above are entirely carried out by the students using Matlab or Python under the supervision of the teachers and represent the most intensive part of the entire course, namely about 20 hours.

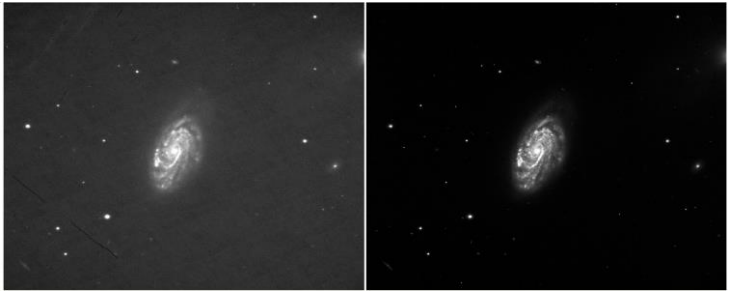

*Figure 2. Raw (left) and calibrated (right) image of the local spiral galaxy NGC 3294 observed at the TNG.*

## 3. Results and discussion

Over our 4-year campaign[2], our students created a database of 75 local spiral galaxies (including those observed at Loiano), 13 of them observed at the TNG. Forty students in total have visited La Palma and 87% of them are now involved in astrophysics studies across multi-levels in academia: 6 Bachelor students, 12 Master students, 15 PhD students (6 of them are based in Firenze) and 2 are currently post-doctoral researchers.

Before the COVID-19 outbreak, the activity in a face-to-face laboratory was characterized by significant interactions between the students and the teachers as well as amongst the groups. Most of the difficulties the students encountered during the laboratories concern coding and handling astronomical files but, to foster reasoning, we favoured a manual that provided only a basic guide to the students, who had to figure out and tailor the examples we provided to their case. This guide was insufficient to carry out the laboratory activities remotely. We had to expand the manual significantly, with a step-by-step reduction for the students to follow, which represented one of the most challenging parts of the course redesign for us as teachers as this choice had pros and cons. The students were faster in carrying out the different steps, but we observed that they were making more basic mistakes, mainly from copying the examples of the lines of code we provided without reasoning on how to tailor the specific line of code for their purposes (for example, modifying the pixel coordinates of the objects for the spectral extraction). Moreover, the students divided in groups were forced to meet in separate virtual rooms, thus the feedback amongst groups was lacking almost completely. The final individual exam consists of an oral examination of the results of the data analysis and a couple of questions on the full course program.

---

[2] 2017, 2018, 2019 and 2021. COVID-19 forced all the observatories to shut down during 2020, whilst the 2021 observing run was done remotely.

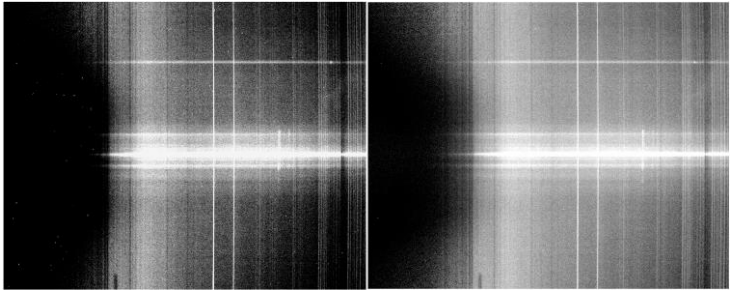

*Figure 3. Raw (left) and calibrated (right) 2D spectral image of the local spiral galaxy NGC 3294 observed at the TNG.*

During the online oral examination, we noticed that a significant fraction of the students who observed remotely did not fully understand the observing procedure done by the support astronomer during the observations, differently from those students that had the chance to be physically present at the telescope. We believe that the observed loss of insights into fundamental observational techniques was mostly driven by the remote observing and the absolute lack of student training on-site. This result is further supported by previous works on this topic, where significant concerns were raised over the limitations that remote observing may introduce (Lockman, 2005; Privon et al. 2009). Since Complementi di Astronomia went from a face-to-face course, with a significant fraction of the lectures done in the lab and observing on-site at the telescope, to 100% virtual format, the interactions among students were considerably reduced, and so was the feedback from the class during the online lectures. An option to ameliorate remote instruction in astronomy could be the use of robotic facilities as performed by The Open University (UK, Brodeur, Kolb, Minocha, Braithwaite, 2014; Kolb, Brodeur, Braithwaite, Minocha, 2018). The Copernico telescope located in the Cima Ekar, near Asiago (Italy) is moving towards this direction, but it is not fully operational in this way yet. Even though the study by Brodeur et al. (2014) shows that practical astronomy can be effectively taught through fully-remote methods, such an approach is still not comparable to on-site observation. As observational astrophysics is at the heart of our course, the COVID-19 pandemic not only affected the overall course logistic but also had an influence on the students' motivation. During the trip to La Palma that lasts around 5 days, the students have also the unique opportunity to visit several cutting-edge telescopes, such as the Gran Telescopio Canarias, the William Herschel Telescope, the Swedish Solar Tower, and the MAGIC (Major Atmospheric Gamma Imaging Cherenkov) telescope. Moreover, most of our students live in the city and thus have never had the chance to watch a real dark sky with the Milky Way, let alone star constellations and the zodiacal light. We believe that, regardless of what will be their path after their university studies, this experience had a profound impact on their future career, forging a stronger scientific and critical mindset in our students.

## 4. Conclusions

With the current pandemic situation, observing on-site is still prohibited, thus we will continue with remote observing for the 2022 academic year. We highlight that this activity is only possible thanks to a private donation and the availability and direct collaboration with the TNG staff, which reserves us a few hours every year. The overall budget required in 2019, for instance, was roughly 15,000 euros for a total of 15 participants (12 students and 3 teachers), which represents a significant expense, unaffordable for a typical Italian university study program in Physics or Astronomy. Therefore, only an external fund can make the entire experience possible. TNG is a cutting-edge telescope, we are thus privileged to have the opportunity to observe at this, significantly oversubscribed, facility. To conclude, the authors wish to thank the Galileo Galilei Fundaciòn and INAF for the allotted observing time, the hospitality and the logistic and scientific support; Dr. Maria Grazia Magini, for the donation to support educational activity and young researchers in memory of the late Stefano Magini the Arcetri Astrophysical Observatory to have offered part of the donation to this activity; and finally the Department of Physics and Astronomy of the University of Firenze for the financial and logistic support to the initiative.